\documentclass{optica-article}

\journal{opticajournal} 

\articletype{Research Article}
\usepackage{lineno}

\begin{document}

\title{Wavefront shaping through a free-form scattering object}

\author{Alfredo Rates,\authormark{1,*} Ad Lagendijk,\authormark{1} Aur\`{e}le J. L. Adam,\authormark{2} Wilbert L. IJzerman,\authormark{3,4} and Willem L. Vos\authormark{1}}

\address{\authormark{1}Complex Photonic Systems (COPS), MESA+ Institute for Nanotechnology, University of Twente, P.O. Box 217, 7500AE Enschede, The Netherlands\\
\authormark{2}Optics Research Group, Department of Imaging Physics, Delft University of Technology, Lorentzweg 1, 2628 CJ Delft, The Netherlands\\
\authormark{3}CASA, Department of Mathematics and Computer Science, Eindhoven University of Technology, PO Box 513, 5600MB Eindhoven, The Netherlands\\
\authormark{4}Signify Research, High Tech Campus 7, 5656AE Eindhoven, The Netherlands\\
}

\email{\authormark{*}a.ratessoriano@utwente.nl}

%
%
\begin{abstract*} 
Wavefront shaping is a technique to study and control light transport inside scattering media. 
Wavefront shaping is considered to be applicable to any complex material, yet in most previous studies, the only sample geometries that are studied are slabs or wave-guides. 
In this paper, we study how macroscopic changes in the sample shape affect light scattering using the wavefront shaping technique. 
Using a flexible scattering material, we optimize the intensity of light in a focusing spot using wavefront shaping and record the optimized pattern, comparing the enhancement for different curvatures and beam radii. 
We validate our hypothesis that wavefront shaping has a similar enhancement regardless of the free-form shape of the sample and thus offers relevant potential for industrial applications. 
We propose a new figure of merit to evaluate the performance of wavefront shaping for different shapes. 
Surprisingly, based on this figure of merit, we observe that for this particular sample, wavefront shaping has a slightly better performance for a free-form shape than for a slab shape. 
\end{abstract*}
%
%

%
\section{Introduction}
%
%

How to transport light efficiently from A to B? This deceivingly simple question is the central challenge in the functionality of daily used devices such as cameras, projectors, lighting systems, and even optically secured bank cards. 
When light travels from point A to B through a transparent medium with macroscopic, optionally curved, surfaces it follows a path that --- following the famous principle of Fermat (1658) --- has an optical length which is an extremum~\cite{Winston2005book}. 
This changes fundamentally when the medium contains microscopic particles that scatter light. 
The technologically relevant question ``what happens when both macroscopic curved interfaces and nanophotonic scattering media appear simultaneously", has remained unaddressed and unanswered yet. 
To date, free-form optics and nanophotonics have developed separately from each other, the former originating from optical engineering and the latter from condensed matter physics. 
The interplay between microscopic scattering and refraction or reflection by macroscopic free-form surfaces offers new, largely unexplored, opportunities and solutions to diverse technological problems. 

The question of how light travels efficiently from A to B is central to many advanced technologies that play important roles in modern society, such as high-precision metrology for integrated nano-circuits in the Internet of Things (IoT) applications and our smartphones, or detectors in earth-observing satellites. 
In modern optics, the question is addressed by custom-designing free-form components – lenses and curved mirrors – that transfer a given distribution of light at the source to a desired distribution at the target in real space, or a solid angle in angle space. 
Over the last decade, free-form optics have been used in the development of versatile, miniature, and efficient devices that appear in daily life~\cite{Winston2005book,fang2013cirp}. 

\begin{figure}[tbp]
    \centering
    \includegraphics[width=0.8\columnwidth]{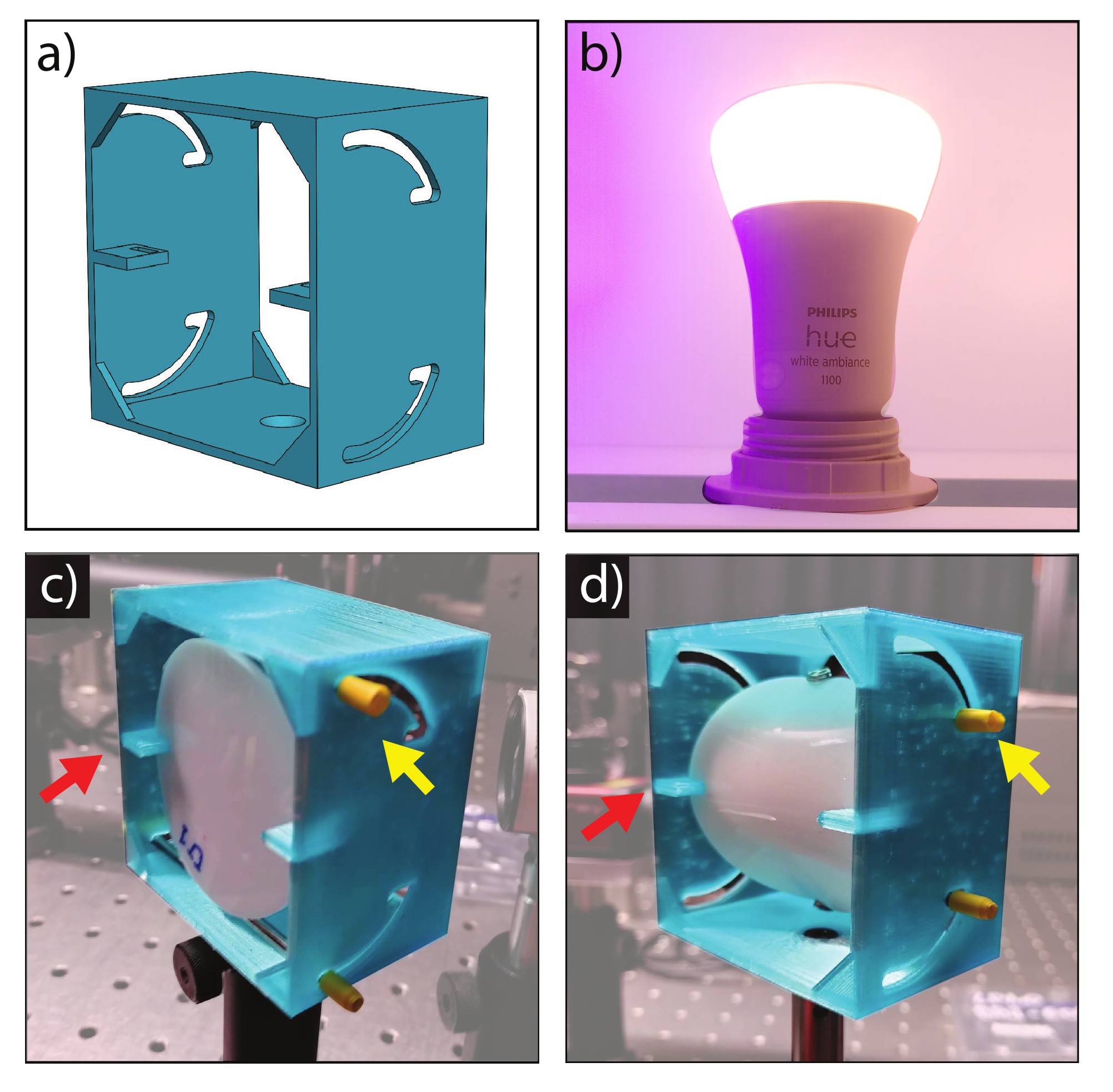}
    \caption{
    Tailored-made holder and sample used for experiments. 
    a) 3D model of the sample holder. 
    b) Commercial smart light bulb, Philips HUE, from which the free-form shape of our sample is inspired. 
    c) The sample in the sample holder in the flat position, and d) the sample in the curved position. 
    The sample is attached horizontally to the holder (red arrow) and vertically to two movable metal bars (yellow arrow). 
    The bars move in the range delimited by the rails on the side of the holder. 
    }
    \label{fig:setup_sample}
\end{figure}

While nanophotonic media that scatter light cannot be understood with current free-form design methodologies, many modern devices greatly benefit from these media. 
For instance, in our homes, offices, and streets, a quiet revolution is taking place as white LEDs are replacing energy-inefficient light sources. 
A white LED consists of a blue diode~\cite{Nakamura2015annderphys}, whose output is converted and diffused in a layer of phosphor particles~\cite{Schubert2006book, Krames2007jdt}. 
The presence of the light-scattering (and also absorbing and re-emitting) phosphor layer is essential to the functionality of a white LED. 
Other applications of nanophotonic media include high-precision metrology tools for nanolithography, the calibration of space instruments for earth observations, and novel (quantum) optically secured information technology for privacy. 
Currently, it is not feasible to describe light-scattering coatings, diffusers, or suspensions with free-form optics. 
Conversely, in nanophotonics, analytic solutions exist only for simple sample geometries such as a sphere, slab, and plane. 
No analytic solutions exist for arbitrary free forms. 
The presence of microscopically structured materials in macroscopic free forms implies a huge difference in scale to which conventional optical models cannot be applied. 
Hence current industry solutions invoke shortcuts, including untested assumptions (\textit{e.g.} material homogeneity or disregard wavelength dependence). 
Today’s lack of knowledge on free-shape scattering optics hampers fast, efficient, and systematic design progress as well as the development of new optical architectures. 

One method extensively used for studying nanophotonic scattering media is wavefront shaping (WFS), where the light propagation through a scattering medium is controlled by interference \cite{Vellekoop2007OptLett, Mosk2012NatPhot, Popoff2014prl, Vellekoop2015optexp, rott2017rmp}. 
These interferences generate a random pattern in the far field, called the speckle pattern. 
Recently, the potential of WFS has been extended to, for instance, time-varying samples~\cite{Nixon2013nature, Horstmeyer2015np, Gigan2022iop}, and periodic samples such as 2D (and 3D) photonic crystals~\cite{Sarma2017apl, Uppu2021prl}. 
The interferences inside a scattering media are considered random. 
But still, when certain properties are slightly changed, a correlation in the speckle pattern persists. 
The phenomenon of persisting correlations is called the memory effect and has been studied by changing the angle of the incident beam, its position on the sample, and its wavelength~\cite{Freund1988prl, Feng1988prl, Judkewitz2015nature, vanBeijnum2011optlett}. 
A \textit{nonlinear} optical memory effect has been recently reported with nonlinear optothermics~\cite{Fleming2019oe}. 
In all these studies, however, the sample's shape has a high symmetry (typically a slab) that is kept constant. 

In most cases to date, WFS has been done on the quintessential scattering sample shape, namely in slabs. 
However, as previously illustrated, real-world applications require samples to have a free-form shape and be finite. 
Although many robustness studies have been made in WFS, as far as we know, no study includes changes in the sample's shape. 
Because of that, the impact of the macroscopic characteristics of the object is not addressed in current theories. 

Here, we present the study of an opaque sample of TiO$_2$ particles suspended in silicone, as shown in Figure~\ref{fig:setup_sample}. 
Exploiting the mechanical flexibility of silicone, we modify the shape of the sample and measure the enhancement of the intensity $\eta$ in a portion of the speckle pattern in angle space, \textit{i.e.}, a solid angle. 
We used a tailored-made sample holder to change both the curvature of the flexible sample and a focusing lens with a linear stage to change the beam radius on the sample. 
When it is curved, the sample has a curvature radius of approximately $R_{\rm c} = 15$~mm. 
The thickness and the radius of curvature of the slab are inspired by the dimensions used in, \textit{e.g.}, the Philips HUE bulb, see Figure~\ref{fig:setup_sample}b). 
In this product, a set of colored LEDs is used to create both colored and white light at will using wireless control. 
To mix the light of these LEDs, the outer bulb plays an instrumental role. 
The typical thickness of the bulb is close to 2~mm, and the radius varies with position, of which we considered $15$~mm a typical value. 
We compare the performance of WFS in a flat and a free-form sample as a function of beam radius. 
With this, we want to draw the attention of the community to study the properties of industrially relevant free-form scattering objects both in theory and experiments.

%
\section{Experimental Methods}
\begin{figure}[tbp]
    \centering
    \includegraphics[width=0.8\columnwidth]{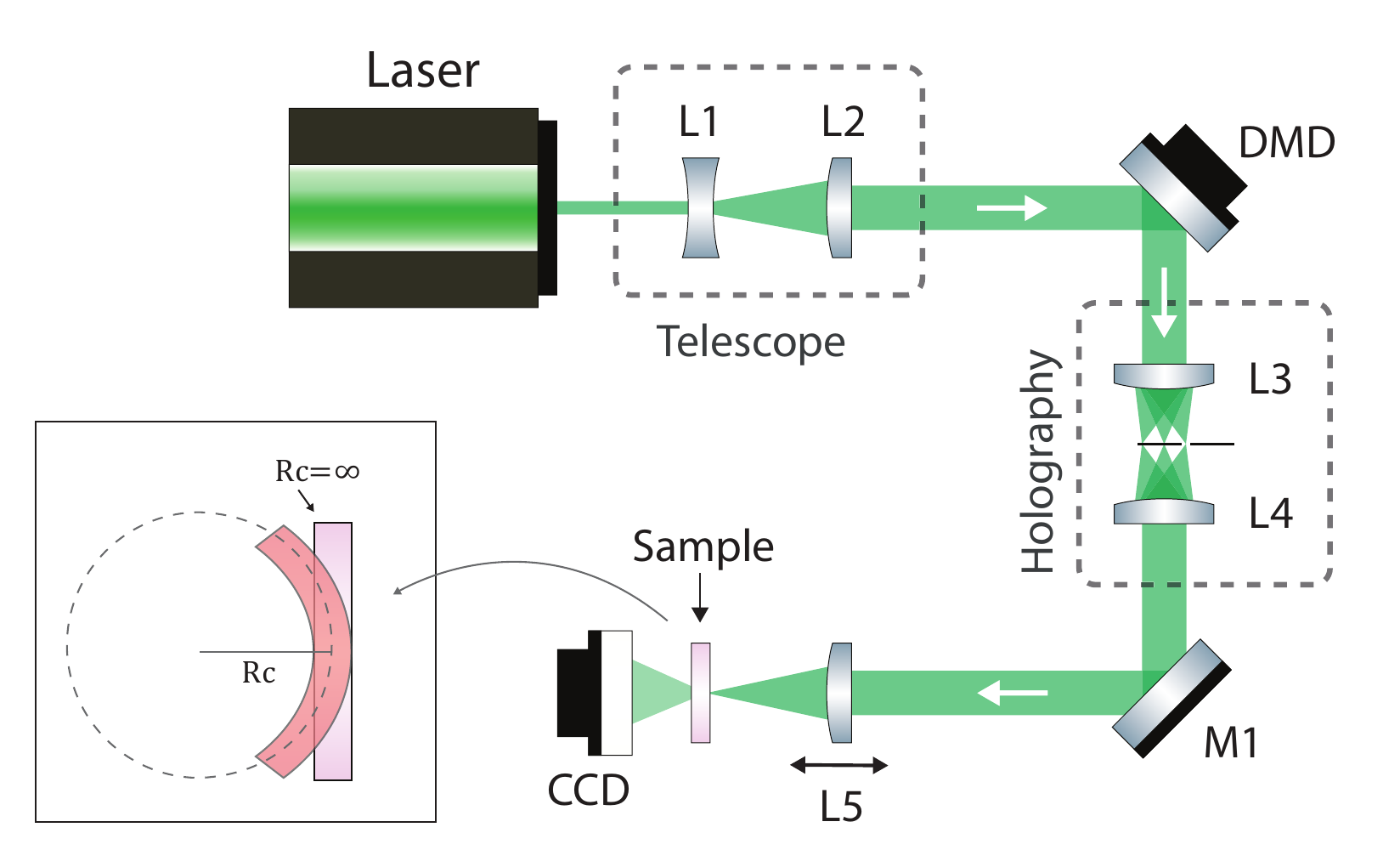}
    \caption{
    Experimental setup for Wavefront Shaping. 
    We use a DMD device combined with the Lee holography method to achieve phase-only modulation. 
    Lens L5 is placed on top of a linear stage, to control the distance between L5 and the sample. 
    We collect the speckle pattern in transmission. 
    Inset: diagram of the curvature of the sample $R_{\rm c}$. 
    The sample is curved in such a way that the position of the center of the sample is constant. 
    The sample is curved using the holder shown in Figure~\ref{fig:setup_sample}. 
    (L: lens; M: mirror).
    }
    \label{fig:setup_experiment}
\end{figure}

%
%
The optical setup is depicted in Figure~\ref{fig:setup_experiment}. 
We use a green laser (Coherent COMPASS 315M-100, $\lambda=532$~nm) as a source, and we modulate the wavefront using a digital micromirror device (DMD, DLP7000 Texas Instruments). 
To convert the binary amplitude modulation into a phase modulation, we use the Lee Holography technique~\cite{Lee1978progrInOpt, Conkey2012optexp} implemented with a 4f system, where we filter the $-1$ order in the Fourier plane with an iris. 
The beam is then focused into the scattering material, placed in a tailored-made holder, and a CCD camera (Guppy PRO F-125b) collects the light transmitted through the sample. 
The focusing lens L5 is placed on top of a linear stage to have control over the position of the sample related to the focal length. 
The information collected from the CCD camera is used to optimize the DMD pattern following the sequential algorithm~\cite{Vellekoop2007OptLett, Vellekoop2008PRL}.

The sample consists of TiO$_2$ particles suspended in silicone with a weight concentration of 0.1 wt\%. 
The sample has a wafer-like shape with a thickness of $d=2$~mm.
To study the free-form sample, we exploit the flexibility of silicone to do measurements when the sample is flat and perpendicular to the incident beam and when the sample is curved. 
To curve the sample in a controlled manner, we designed a special sample holder, shown in Figure~\ref{fig:setup_sample}. 
The sample is attached to the holder from the side with clamps and on the top and bottom to sliding metallic bars.
These fixed points are highlighted in Figure~\ref{fig:setup_sample}c) and Figure~\ref{fig:setup_sample}d) with yellow and red arrows, respectively. 
When the bars are close to the front, the sample is completely flat, and when the bars are fixed close to the back, the sample is curved. 
With this holder, we control the curvature of the sample during scattering and wavefront shaping experiments while other physical parameters are constant. 

%
\section{Enhancement in curved objects}
%
%

A relevant parameter to study in WFS experiments is the intensity enhancement $\eta$. 
The enhancement is defined as

\begin{align}\label{eq:eta}
    \eta \equiv \frac{I_{\rm opt}}{\langle I_{\rm 0}\rangle},
\end{align}
\\
where $I_{\rm opt}$ is the intensity at a solid angle after wavefront optimization and $\langle I_{\rm 0}\rangle$ is the ensemble average intensity of the pattern before optimization~\cite{Aulbach2012optexp}. 
The ensemble average intensity is calculated by averaging the total counts on the detector at the target position over multiple random wavefronts. 
The enhancement increases when the number of segments $N$ increases. 
A \textit{segment} is a finite area of the incoming modulated wavefront that has the same phase value. 

\begin{figure}[tbp]
    \centering
    \includegraphics[width=0.9\columnwidth]{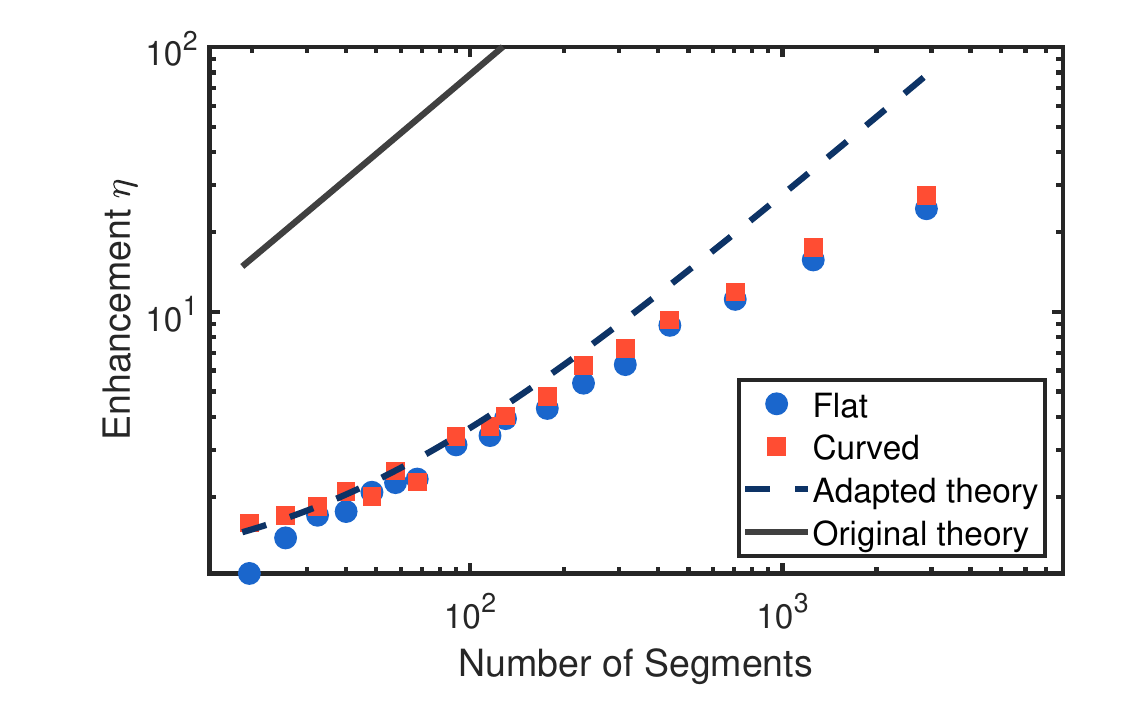}
    \caption{
    Enhancement versus the number of active modulating segments. 
    The illuminating spot has a diameter $2r_{\rm b}=4.5mm$. 
    The blue circles represent the enhancement when the sample is flat and the red squares are for the curved sample. 
    Each symbol is an average from three independent repetitions of the optimization. 
    The dashed curve is the theory including correcting for the variable speckle size, and the solid curve is the theoretical maximum without correction. 
    Wavefront shaping is as effective in optimizing the intensity for a curved sample as for a flat sample. 
    }
    \label{fig:flatVcurve}
\end{figure}

We compare a free-form sample with a slab-like shape by measuring the enhancement $\eta$ for a different number of segments $N_{\rm s}$. 
In Figure~\ref{fig:flatVcurve}, we plot $\eta(N_{\rm s})$ for both a flat and a curved sample. 
Each symbol is composed of the maximum enhancement obtained with the corresponding number of segments, averaged over three iterations of the optimization loop. 
Every iteration uses a flat incident wavefront as the initial value, meaning each optimization is independent of the others but starts from the same initial condition.
Figure~\ref{fig:flatVcurve} also includes two curves corresponding to theoretical limits. 
We see that the maximum theoretical limit is considerably above the experimental data. 
The discrepancy between the maximum limit and experimental data is because the limit does not take into account variations in the speckle size. 
In section~\ref{sec:theo} we will elaborate an extension of the theoretical limit, which yields values much closer to the experimental data. 

Figure~\ref{fig:flatVcurve} shows that both sample shapes give the same trend in enhancement, indicating that the performance of the wavefront shaping technique is the same, regardless of the macroscopic properties of the object. 
This performance persistence agrees with our hypothesis, and with the current theory of wavefront shaping~\cite{Freund1990PhysicaA, VanRossum1999rmp}. 
If we interpret the scattering sample with a random transmission matrix, regardless of the value of each coefficient, the efficiency of the wavefront optimization is unchanged. 
A possible effect that may change the total enhancement is when the number of open channels changes. 
If the number of open channels is less than the number of controlled segments, the enhancement saturates, such that an increased number of segments does not further increase the enhancement. 
Our data shows no indication of a saturation, meaning that the number of open channels is apparently unchanged. 

\begin{figure}[tbp]
    \centering
    \includegraphics[width=0.9\columnwidth]{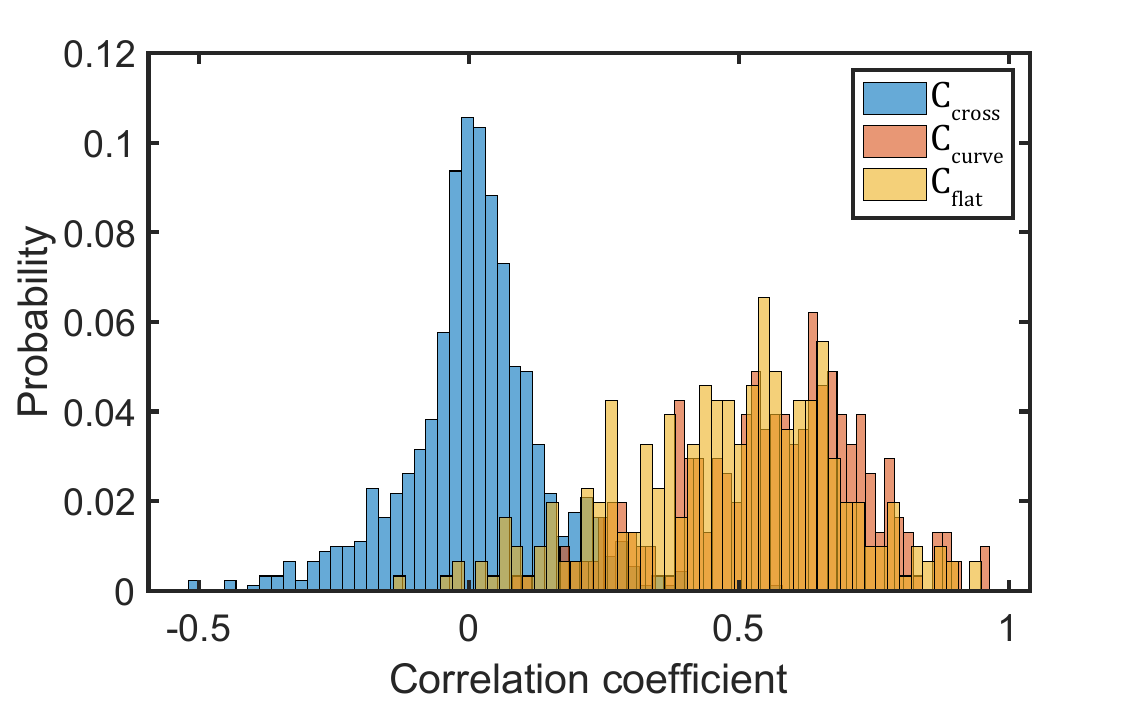}
    \caption{
    Correlation distribution between optimized phase patterns. 
    The blue distribution corresponds to the correlation between patterns from different shapes, while the orange and yellow distribution corresponds to the correlation between the same sample's shape (curve and flat, respectively). 
    }
    \label{fig:histo}
\end{figure}

An alternative interpretation of these results is that the curvature is simply not affecting the light transport inside the material or that it only affects the border of the optimized wavefront with a curvature effect. 
To test this new hypothesis, we calculate the correlation between the optimized wavefront of different measurements. 
For each number of segments, we have a total of 6 measurements, three from a curved sample and three from a flat sample. 
Between measurements with the same geometry, all the conditions are kept the same, and only measurement noise could cause any discrepancy.
We take the optimized wavefront for each measurement, and we calculate the Pearson correlation coefficient between the 6 cases.
We use the SciKit python library to calculate the correlation, which is defined as~\cite{Pedregosa2011online}    

\begin{align}\label{eq:pearsonA}
    C_{\rm P} \equiv \frac{\sum_{\rm i,j}(x_{\rm i,j}-\overline{x})(y_{\rm i,j}-\overline{y})}{\sqrt{\sum_{\rm i,j}(x_{\rm i,j}-\overline{x})^2 \sum_{\rm i,j}(y_{\rm i,j}-\overline{y})^2}},
\end{align}
\\
with $x_{\rm i,j}$ the phase value of segment $(i,j)$ of the first wavefront, $y_{\rm i,j}$ the phase value of segment $(i,j)$ of the second wavefront, and $\overline{x}$,$\overline{y}$ the average segment values of the first and second wavefront, respectively. 

We separate the correlations of wavefronts into three classes: the cross-correlations between wavefronts for a flat and a curved sample $C_{\rm cross}$, the correlation between wavefronts for a flat shape only $C_{\rm flat}$, and the correlation between wavefronts for a curved shape only $C_{\rm curve}$.
For the cases where the sample's shape is not changed ($C_{\rm flat}$ and $C_{\rm curve}$), we expect the optimized wavefronts to converge to a similar configuration, therefore a high mutual correlation, regardless of the hypothesis we mentioned before. 

If the macroscopic shape does affect the light transport when we change the sample's shape, we expect the optimized wavefront to converge to a new configuration, and hence $C_{\rm cross}$ to be small. 
On the other hand, if the macroscopic changes do not affect the light transport, then we expect $C_{\rm cross}$ to remain constant regardless of the sample's shape.

The probability distribution of the correlations is shown in Fig,~\ref{fig:histo}. 
We see that $C_{\rm cross}$ is centered at 0 and extends from about $-0.3 < C_{\rm cross} < +0.3$. 
In contrast, both $C_{\rm flat}$ and $C_{\rm curve}$ have higher and mostly positive correlations, centered near $0.5$ and extending from $0$ to $1$. 
From these differences in correlations, we conclude that the macroscopic shape of the object affects the light transport inside; apparently, when the macroscopic shape changes, the microstructure of the media changes simultaneously (presumably the scatterers change position relative to each other). 
Thus, to control the scattered light in a free-form shape, we need a new optimized wavefront compared to wavefront shaping on the flat shape. 

In the field of light scattering, the \textit{memory effect} is widely known~\cite{Freund1988prl, VanRossum1999rmp, Judkewitz2015nature, Adhikary2021thesis}. 
The memory effect shows that when the light transport inside the scattering media is slightly changed, the speckle (and thus the optimized wavefront) are correlated with the previous one. 
As both speckle patterns are correlated, the same optimized wavefront can be used to enhance the intensity. 
From our observation that curved wavefronts are nearly uncorrelated with flat wavefronts, we conclude that the change in curvature is beyond the range of the memory effect. 
In previous studies, the light transport is changed by displacing the scattering media, tilting the incident beam, or changing the wavelength of the source. 
We thus believe our work is the first systematic study of the free-form memory effect, \textit{i.e.}, the memory effect when changing the shape and form of the scattering media. 

In theory, we expect the distributions of $C_{\rm flat}$ and $C_{\rm curve}$ to tend to 1, while they are centered at 0.5. 
We attribute this to the fact that the DMD is not evenly illuminated because the laser beam has a Gaussian profile.
Thus, the segments at the edges do not affect the optimization significantly and hence the $(x_{\rm i,j},y_{\rm i,j})$ from Eq.~\ref{eq:pearsonA} are different between one iteration and the other, thereby decreasing the correlation. 
We confirm this by looking at the phase difference between individual pixels of the optimized patterns, where only the center has a near-zero difference. 

From a geometrical-optics perspective, one might argue that even in the absence of scatterers, the curvature of the sample could sufficiently impact the wavefront to induce decorrelation of the optimized wavefronts. 
If this were indeed the case, one would anticipate that any decorrelation induced by the geometry would exhibit a pattern consistent with the curvature, \textit{i.e.}, the central region of the wavefront being less affected than its edges.
However, upon closer examination of each individual segment, we observe that the decorrelation is due to random changes in the complete wavefront, rather than following the expected curvature-related trend. 
Thus, we dismiss this possibility. 

\begin{figure}[tbp]
    \centering
    \includegraphics[width=\columnwidth]{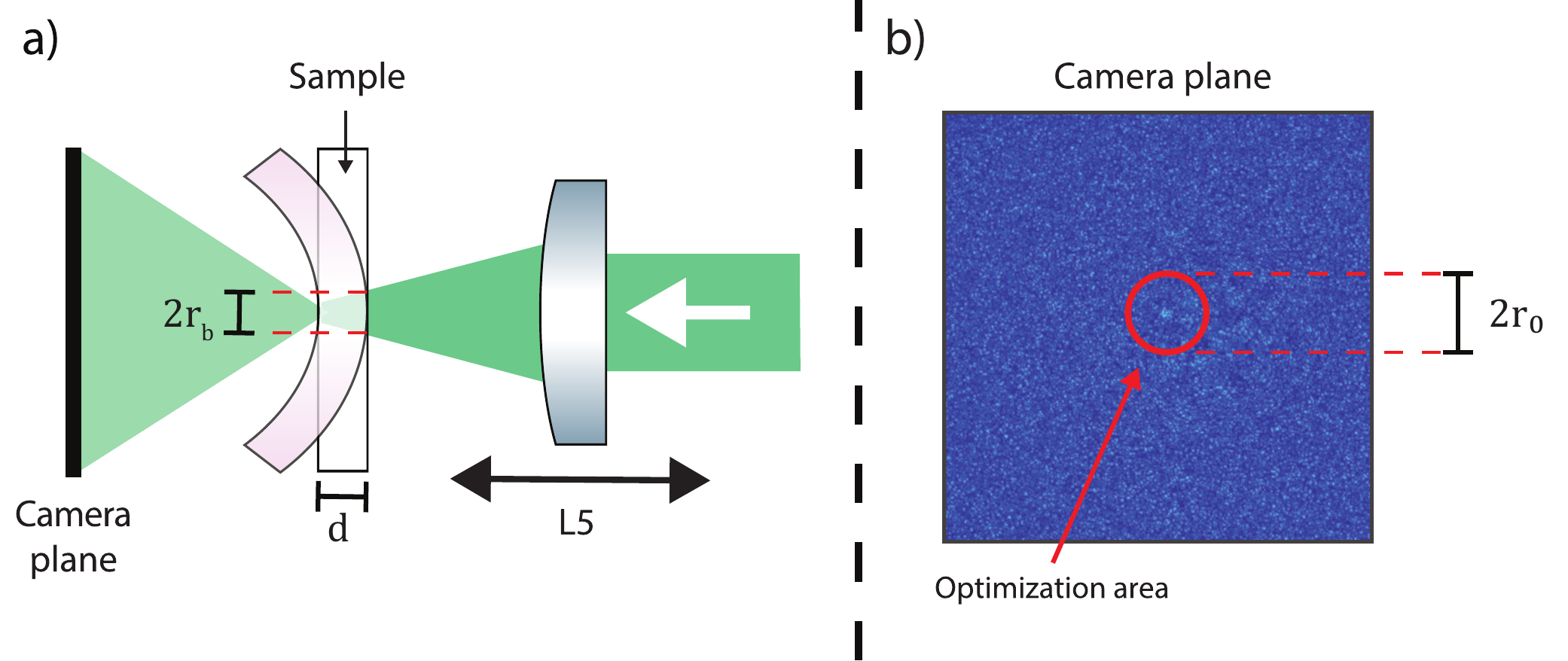}
    \caption{
    a) Cartoon diagram of the illumination area, defined by its radius $r_{\rm b}$. 
    This depends on the distance between lens L5 and the sample. 
    The curvature of the diagram is exaggerated for a didactic purpose. 
    b) Diagram of the optimization area on the camera, defined by its radius $r_{\rm o}$. 
    This depends on the speckle size at the detector.
    To maintain a standard comparison, we keep $r_{\rm o}$ constant in our experiments.
    }
    \label{fig:radii}
\end{figure}

%
\section{Enhancement in presence of variable speckle size}\label{sec:theo}
%
%

To estimate the maximum enhancement possible $\eta_0$ in WFS, one generally invokes wave-guide approximations typical of random matrix theory~\cite{Freund1990PhysicaA,wigner1951book,guhr1998pr,derMolen2007prl} and simplify the scattering medium as a set of scattering channels. 
For this study, We will focus on the intensity enhancement of a single speckle spot as the figure of merit (other types of optimizations are, \textit{e.g.}, enhance the total transmission of the system). 
The maximum enhancement $\eta_0$ is obtained when all the channels are optimized. 
$\eta_0$ expressed as~\cite{Vellekoop2007OptLett} 

\begin{align}\label{eq:theo_enhancement}
    \eta_{0} = \frac{\pi}{4} (N_{\rm s}-1) + 1,
\end{align}
\\
where $N_{\rm s}$ is the total number of segments used in the modulator. 
This equation is valid for phase-only modulation and does not consider the saturation point when $N_{\rm s}$ is equal to the number of transmission channels inside the sample $N_{\rm c}$. 

To compare $\eta_0$ with our experiments, we need to distinguish between the beam radius $r_{\rm b}$ and the optimization radius $r_{\rm o}$. 
This difference is illustrated in Figure~\ref{fig:radii}, where $r_{\rm b}$ is the radius of the illumination area at the sample surface that depends on the focusing lens L5 and the position of the sample. 
Conversely, $r_{\rm o}$ is the radius of the area of the speckle pattern that we observe at the detector, and it is proportional to the speckle size of the speckle pattern. 
Because we have a lens-less optical system, this area represents a solid angle. 

\begin{figure}[tbp]
  \centering
  \includegraphics[width=\linewidth]{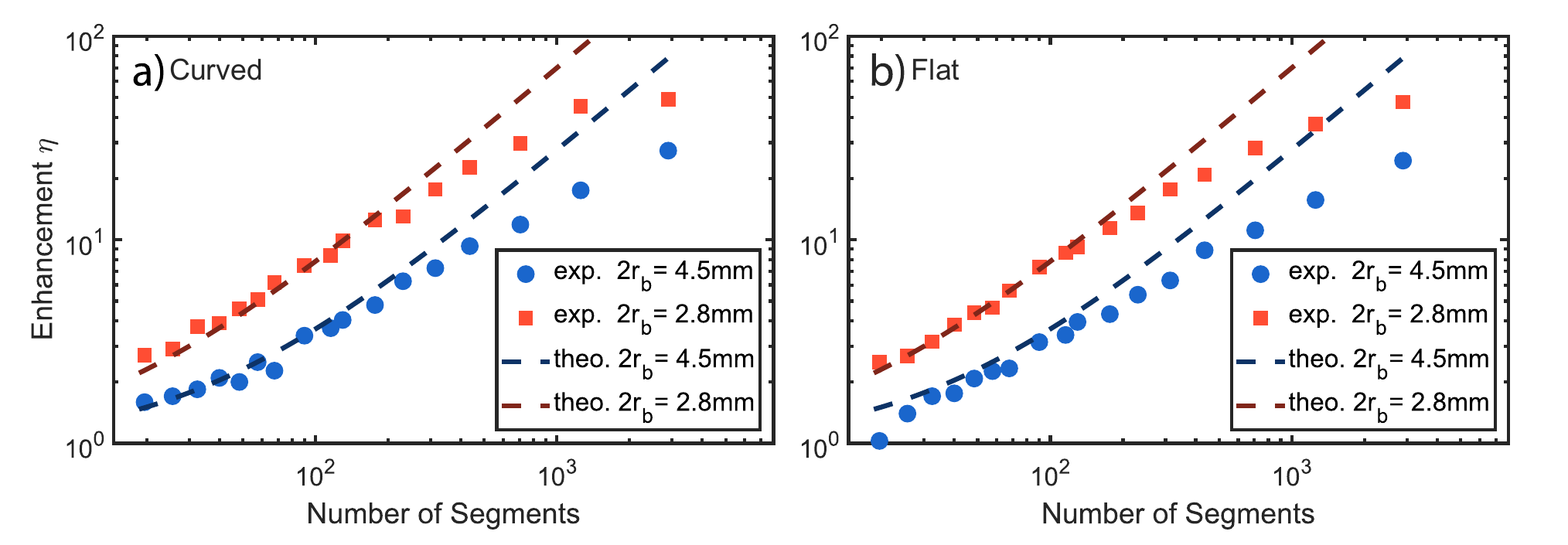}  
    \caption{
        Enhancement versus the number of active modulating segments for a) flat shape and b) curve shape. 
        On each plot, we show the enhancement for the maximum ($2r_{\rm b}=4.5mm$, blue circles) and minimum ($2r_{\rm b}=3.8mm$, red squares) illumination radius. 
        The dashed lines show the theoretical enhancement corrected based on $r_{\rm b}$.
    }
  \label{fig:Radii}
\end{figure}

Although the optimization process does not depend on $r_{\rm b}$ directly, it depends on $r_{\rm o}$: to get the maximum enhancement, the optimization area must have the same size as a single speckle $r_{ \rm o} = r_{ \rm s}$. 
The size of this single speckle in the detector plane depends on the design of the experiment, particularly if we use or not a lens at the detector. 
For our measurements, we use a lens-less system, hence the size of a single speckle is a solid angle. 
At a distance $d$, the speckle is projected with a radius equal to~\cite{Fujiwara2021iop}:
\begin{align}\label{eq:sizeSpeckle}
    r_{ \rm s} = \frac{ 2.44\cdot \lambda }{4 r_{\rm b} } d,
\end{align}
\\
with $\lambda$ the wavelength of the incident light and $d$ the thickness of the sample. 
We infer from Eq.~\ref{eq:sizeSpeckle} that $r_{\rm o}$ and $r_{\rm b}$ are inversely related, which means the optimization process depends on both parameters. 

In Ref.~\cite{Ojambati2016pra}, Ojambati and co-workers studied how the maximum enhancement changes when changing the optimization radius $r_{ \rm o}$, thus $\eta_0' = \eta_0'(r_{ \rm o})$. 
This dependency is expressed as~\cite{Ojambati2016pra}

\begin{align}\label{eq:correcterOptEnh}
    \eta_{0}'(r_{ \rm o}) = \frac{K}{r_{ \rm o}^2} + 1,
\end{align}
\\
where the correction factor $K$ depends on the number of segments $N_{\rm s}$ and on the speckle size.
The theoretical limit $\eta_0$ is reached when $r_{ \rm o}$ approaches the speckle size $r_{ \rm s}(r_{ \rm b})$, $\lim_{r_{ \rm o} \to r_{ \rm s}(r_{ \rm b})} \eta_{0}'(r_{ \rm o}) = \eta_0$. 
If we take the limiting value of $r_{ \rm o}=r_{ \rm s}(r_{ \rm b})$ in Eq.~\ref{eq:correcterOptEnh} and we replace it with Eq.~\ref{eq:sizeSpeckle}, and we take the limiting value of  $\eta_{0}'(r_{ \rm o})=\eta_0$ and we replace it with Eq.~\ref{eq:theo_enhancement}, we obtain an extended correction factor $K(N_{\rm s}, r_{ \rm b})$:

\begin{align}\label{eq:correctionFactor}
    K(N_{\rm s}, r_{ \rm b}) = \frac{ 2.44^2 \pi \lambda^2 }{ 4^3 r_{\rm b}^2 } d^2 (N_{\rm s}-1).
\end{align}
\\
Finally, the theoretical enhancement corrected for optimization and beam areas is given by 

\begin{align}\label{eq:correcterOptEnhFinal}
    \eta_{0}'(r_{ \rm o}, r_{ \rm b}) = \frac{ 2.44^2 \pi \lambda^2 }{ 4^3 r_{\rm b}^2 r_{ \rm o}^2 } d^2 (N_{\rm s}-1) + 1.
\end{align}
\\
In Eq.~\ref{eq:correcterOptEnhFinal} we see that the enhancement is inversely proportional to $r_{\rm b}^2$ and $r_{\rm o}^2$, and directly proportional to $d^2$. 
Both dependencies are in accordance with usual WFS experiments, where one focuses the incident beam to a narrow spot and reduces the optimization area to a single speckle, while placing the detector far from the sample to increase the size of the speckle on the detector. 
In our experiments $r_{\rm o}$ is fixed, and we compare the measurements while varying $r_{ \rm b}$.

%
\section{Effect of sample curvature}\label{sec:curvature}
%
%

We study the impact of the ratio between the beam radius $r_{\rm b}$ and the radius of curvature of the sample $R_{\rm c}$.
We vary the radius $r_{\rm b}$ in a range between 1.9~mm and 2.3~mm (hence diameters $2r_{\rm b} = 3.8$ to $4.6$~mm) by changing the position of the focusing lens L5 (see Figure~\ref{fig:setup_experiment}).
We measure the enhancement versus number of segments $N_{\rm s}$ (as in Fig~\ref{fig:flatVcurve}) for six different beam radii $r_{\rm b}$ and both flat and curved sample's shapes.

Figure~\ref{fig:Radii} shows the enhancements for the lowest and highest $r_{\rm b}$ for both shapes, along with the maximum theoretical limit for each case, calculated from Eq.~\ref{eq:correcterOptEnhFinal}. 
We see that the enhancement decreases when increasing the beam radius, and so does the theoretical limit, in accordance with our theory (see Eq.~\ref{eq:correcterOptEnhFinal}). 
When increasing $r_{\rm b}$, the speckle size is smaller. 
If $r_{\rm o}$ is constant, there are more speckles inside the optimization area, thus the maximum enhancement is lower. 

Finally, we compare the enhancement of a curved and a flat sample for different $r_{\rm b}$. 
For this comparison, we propose the following figure of merit $F$, defined as:

\begin{align}\label{eq:FoM}
    F \equiv \langle \frac{\eta_{\rm curve}}{\eta_{\rm flat}} \rangle _N,
\end{align}
\\
with $\eta_{\rm curve}$ the enhancement for the curve shape, $\eta_{\rm flat}$ the enhancement for the flat shape, and $\langle . \rangle_N$ the \textit{geometric mean} over the number of segments $N$.
This figure of merit is applicable regardless of the shift shown in Figure~\ref{fig:Radii} because we expect this shift to occur both in the flat and the curved shape.

\begin{figure}[tbp]
    \centering
    \includegraphics[width=0.9\columnwidth]{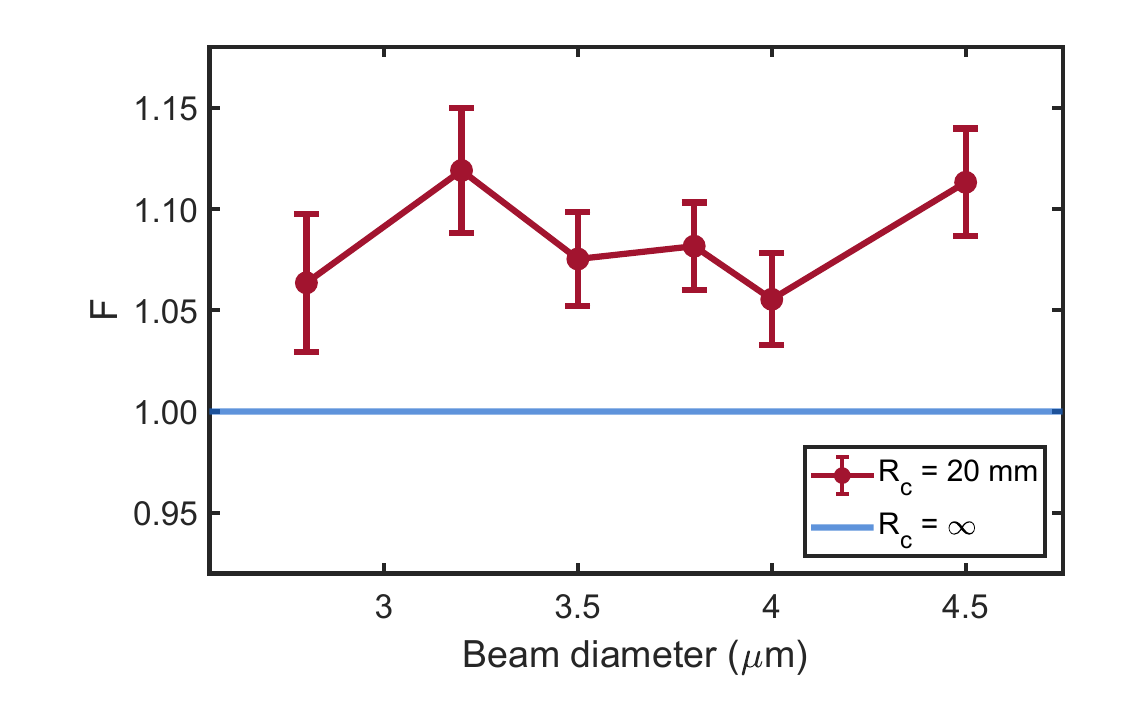}
    \caption{
    Figure of merit $F$ (Eq.~\ref{eq:FoM}) versus beam diameter $2r_{\rm b}$. 
    $F$ is based on comparing a curved sample with $R_{\rm c}=20$ mm to a flat sample ($R_{\rm c}=\infty$). 
    The light blue line at $F = 1$ corresponds to the case when the curvature is not changed.
    }
    \label{fig:FoM}
\end{figure}

We show $F$ in Figure~\ref{fig:FoM} for different beam radii. 
We see that $F$ is constant at about $F=1.08$ and it is independent of $r_{\rm b}$. 
We note that all data exceed $F = 1$, showing that on average, we achieve a larger enhancement with a curved shape compared to a common flat slab-like sample shape. 
This result challenges the current theoretical framework of wavefront shaping because no fundamental scattering property was changed except the macroscopic sample's shape and the position of the lens. 
Based on this figure of merit, the enhancement difference is statistically significant, as all data have $F > 1$ by a difference larger than the confidence interval, with a highly significant p-value ($p \approx 2.8 \times 10^{-8}$).

%
%
\section{Conclusions and outlook}
%
%

In this paper, we have studied how the intensity enhancement obtained using wavefront shaping changes when we change the macroscopic properties of the object, in particular the sample's shape. 
We compare the case of two macroscopic shapes, a free-form curved sample and a flat sample with a slab shape. 
We designed a tailored-made sample holder to control the curvature of a flexible sample and a focusing lens with a linear stage to control the beam radius on the sample. 

We observe a low correlation between the optimal wavefronts for different macroscopic shapes, suggesting that the light transport inside the scattering media changes on a mesoscopic level. 
Meanwhile, when optimizing the intensity using the wavefront shaping technique, we reach a similar enhancement for a curved sample as for a flat sample. 
This performance persistence agrees with our hypothesis; regardless of the changes in light transport, the number of open channels is unchanged, thus the maximum enhancement is not affected by these changes. 
Our results confirm the notion that wavefront shaping is applicable regardless of the form of the object. 

We derive an extension of the theory for maximum enhancement based on the beam and optimization radius. 
Our experiments agree well with our theoretical prediction, and our theory explains the trend for both a curved and flat sample. 
Furthermore, we propose a figure of merit to compare the enhancement between different sample shapes. 
Based on this figure of merit, we observe an increase in the total enhancement by about $10\%$ for every beam radius. 
This increase is not explained with current theories for wavefront shaping based on slab shapes only, and thus calls for new theory. 

With this paper, we aim to encourage the community to consider free-form objects both in experiments and theory of light scattering, as they play a major role in industrial applications. 
Next steps in this project include studying speckle decorrelations in real-time while changing the shape of the sample and studying samples with various densities of scatterers. 
Probing speckle decorrelations in real-time will yield the curvature range of the free-form memory effect for a material, which is relevant for applications. 
For instance, when a light scattering device heats up while being used in a real-world environment (for instance, in semiconductor metrology, in lighting, and in earth observing satellite optics), its shape may change, and hence it is relevant to characterize (and possibly enlarge) the temperature operation range. 

\begin{backmatter}
\bmsection{Funding}
Nederlandse Organisatie voor Wetenschappelijk Onderzoek (NWO-TTW program P15-36). 

\bmsection{Acknowledgments}
We thank Cornelis Harteveld for expert technical support and sample preparation, and Mario Vretenar for the 3D printing of the sample holder. 
We acknowledge the support of all participants of the TTW Perspectief program ``Free-form scattering optics" (FFSO) in collaboration with TU Delft, TU Eindhoven, and industry users ASML, Demcon, Lumileds, Schott, Signify, and TNO. 
This work was also supported by the MESA+ Institute section Applied Nanophotonics (ANP).

\bmsection{Disclosures}
WLV is a shareholder in QuiX Quantum B.V.

\bmsection{Data availability} 
The data used for this publication are publicly available in the Zenodo database in Ref.~\cite{zenodoDatabase}. 
\end{backmatter}

%
%
\bibliography{References.bib}
%
%

\end{document}